# Ultrafast crystallization and sintering of $Li_{1.5}Al_{0.5}Ge_{1.5}(PO_4)_3$ glass and its impact on ion conduction

Antonino Curcio[a], Antonio Gianfranco Sabato[b], Marc Nuñez Eroles[b], Juan Carlos Gonzalez-Rosillo[b], Alex Morata[b], Albert Tarancón[b,c,*], and Francesco Ciucci[a,d,e,f,*]

[a] Department of Mechanical and Aerospace Engineering, The Hong Kong University of Science and Technology, Hong Kong SAR, China

[b] Department of Advanced Materials for Energy, Catalonia Institute for Energy Research (IREC), Jardins de Les Dones de Negre 1, 08930, Sant Adrià del Besòs, Barcelona, Spain

[c] Catalan Institution for Research and Advanced Studies (ICREA), Passeig Lluís Companys 23, 08010, Barcelona, Spain

[d] Department of Chemical and Biological Engineering, The Hong Kong University of Science and Technology, Hong Kong SAR, China

[e] HKUST Shenzhen-Hong Kong Collaborative Innovation Research Institute, Shenzhen, China

[f] HKUST Energy Institute, The Hong Kong University of Science and Technology, Hong Kong SAR, China

* Corresponding authors: atarancon@irec.cat, francesco.ciucci@ust.hk

**Abstract**

$Li_{1.5}Al_{0.5}Ge_{1.5}(PO_4)_3$ (LAGP) is among the most promising solid electrolytes for the next generation's all-solid-state lithium batteries. However, preparing LAGP electrolytes is time- and energy-intensive. In this work, LAGP glassy powders were sintered and crystallized in 180 seconds by ultrafast high-temperature sintering (UHS) under conditions attractive for continuous industrial processes (*i.e.,* ambient pressure and atmosphere). The fast heating rates characteristic of UHS significantly delay crystallization, potentially decoupling crystallization and sintering. Furthermore, EIS characterizations reveal that LAGP sintered and crystallized by UHS has an ionic conductivity of $1.15 \times 10^{-4}$ S/cm, slightly lower than conventionally annealed samples ($1.75 \times 10^{-4}$ S/cm). The lower conductivity can be attributed to poorer inter-grain contact. To overcome this issue, additives such as $B_2O_3$ and $Li_3BO_3$ are used, resulting in ~2 and ~5 times higher grain boundary conductivity for LAGP+1%wt $B_2O_3$ and LAGP+1%wt $Li_3BO_3$, respectively, compared to LAGP. Overall, this work provides insights into unraveling the impact of UHS sintering on the LAGP $Li^+$ conduction mechanism.

# 1 Introduction

Lithium-ion batteries are a vital technology in the electrification of the transportation and sustainable energy sectors.[1] However, lithium-ion batteries currently suffer from serious safety risks due to the thermal instability and flammability of liquid electrolytes.[2-5] As solid-state electrolytes do not leak, are generally non-flammable, and have better thermal stability than liquid electrolytes, using them can alleviate the safety issues of conventional lithium-ion batteries.[6] To transition to solid-state batteries, an ideal solid-state electrolyte should have $Li^+$ conductivity above $10^{-4}$ S/cm at room temperature.[3] $Li^+$ conducting oxides with a NASICON-type structure are among the most promising solid-state electrolytes as their ionic conductivity is in the $10^{-4}$-$10^{-3}$ S/cm range.[7-9] However, sintering these solid-state electrolytes is a time- and energy-consuming process, requiring long thermal treatments (tens of hours) and high temperatures (up to 1200 °C).[10]

$Li_{1.5}Al_{0.5}Ge_{1.5}(PO_4)_3$ (LAGP), a widely used NASICON-type solid-state electrolyte, is typically produced on an industrial scale by a melt-quenching technique in glass form. Then, LAGP is crystallized with further heat treatment at T>600 °C.[8, 11-12] To achieve a solid electrolyte, crystalline LAGP powders are sintered at high temperatures (800-900 °C for 6 to 12 hours).[11, 13-16] However, crystallization and sintering are lengthy processes occurring at high temperatures, which may cause Li loss leading to the precipitation of secondary phases.[13, 17] Furthermore, the high energy requirements of sintering contribute to a significant portion of the solid-state battery preparation costs.[18-19]

Recently, several rapid techniques for ceramic sintering characterized by fast heating rates and short processing times have been explored, including cold sintering,[20-22] field-assisted sintering technique,[23] flash sintering,[24-25] spark plasma sintering,[26] and ultrafast high-temperature sintering (UHS).[27-29] Among these, UHS is particularly attractive because it allows the preparation of materials within just a few seconds or minutes, and its setup is inexpensive and scalable.[27] Furthermore, a UHS setup allows for atmospheric flexibility and can be easily adjusted to operate in different atmospheres, including Ar, 5% $H_2/N_2$, or air.[30-31] Although UHS has shown great promise for the rapid processing of functional materials, further studies are needed to investigate the opportunities of this technique.

For the first time, LAGP was sintered and crystallized by UHS directly from glass powders in a single step taking just 180 seconds. Further, the strategy represents a significant advance as LAGP pellets are typically produced from crystalline LAGP using either conventional methods or UHS.[11, 13-16, 32] It should be noted that thanks to the short processing time of UHS, Li mass loss and the related formation of unwanted secondary phases were suppressed.[27] The influence of UHS annealing on the sintering and crystallization processes of LAGP was also investigated, as these are sensitive to the heating rate.[33] Furthermore, to document the impact of UHS on the solid electrolyte microstructure and $Li^+$ transport, the physical and electrochemical properties of samples prepared using UHS and conventional methods were compared. Electrochemical characterizations revealed that samples prepared by UHS from bare glass have high grain boundary resistance, likely due to poor inter-grain contact. Therefore, two additives, namely $B_2O_3$ and $Li_3BO_3$, were used to alleviate this issue. During UHS treatment, due to their low melting point (~450 °C and ~700 °C for $B_2O_3$ and $Li_3BO_3$, respectively), these additives are in a liquid phase, which allows an increase in inter-grain contacts and the enhancement of grain boundary $Li^+$ conduction.[3, 34-39]

This article (a schematic representation of its content is given in Figure 1) contributes to unraveling the impact of UHS on the physical properties and $Li^+$ conduction of crystalline LAGP obtained from LAGP glassy powders. These processing conditions (*i.e.* short processing time and ambient atmosphere) are conducive to the large-scale production of solid-state batteries.[40-41] In fact, applying UHS in place of furnace sintering, which has a high impact on energy demand and costs,[19] can represent a viable and less time and energy-intensive alternative.

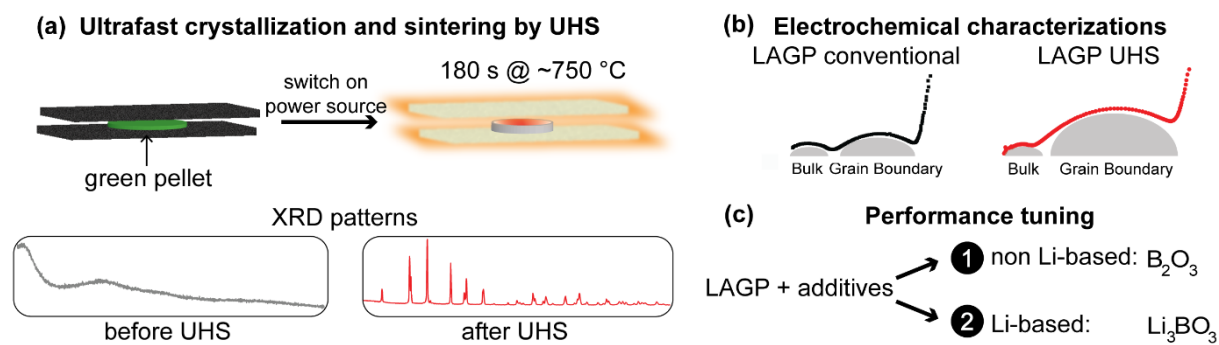

Figure 1. Schematic representation of this work; (a) UHS for ultrafast sintering and crystallization of glassy LAGP. (b) Electrochemical characterization of samples sintered and crystallized by conventional sintering in furnace and UHS, highlighting the significant grain-boundary contribution to the total resistance in LAGP UHS compared to LAGP conventional. (c) The high grain boundary resistance of LAGP UHS is reduced through the use of the $B_2O_3$ and $Li_3BO_3$ additives, resulting in enhanced $Li^+$ conductivity.

## 2 Experimental

### 2.1 Sintering of materials

Commercial LAGP amorphous powders (median particle size of 5.47 µm) were purchased from Toshima Manufacturing, Japan. LAGP pelletized powders were either sintered by a conventional heat treatment in a furnace or by UHS. Die-pressed pellets of LAGP glass were sintered in a muffle furnace at 750 °C for 12 hours. These conditions were applied because they have been reported to yield high density (~80%) and $Li^+$ conductivity (~2×10$^{-4}$ S/cm at room temperature) LAGP pellets.[42] Sintering temperatures above 850-900 °C of LAGP glass led to the formation of macropores and pellet deformation,[42-43] which we also observed (Figure S1). As the ramping and cooling rates were set to 5°C/min, the sintering process required more than 16 hours. When performing UHS, the die-pressed pellets of amorphous LAGP were inserted between two graphite felt strips (~4.8 mm in thickness, ~100 mm in length, and ~25 mm in width, AvCarb®, USA) and sintered in ambient air. The graphite strips were connected to a programmable DC power supply (EA-PSI 9080-60T, Elektro-Automatik, Germany), and a current of ~19 A was applied for 180 seconds. An infrared thermometer (RS Pro RS-9862S, RS components, UK) and a type-K thermocouple indicated that a temperature of ~750 °C was obtained at the center of carbon strips, where the sample was placed. To assess the effect of additives, the mixture of LAGP glass + 1% wt of $B_2O_3$ (Merck Sigma-Aldrich, USA) or $Li_3BO_3$ (Toshima Manufacturing, Japan) was die-pressed and annealed by conventional and UHS methods described above. Preliminary experimental results (Figure S2) showed that additive amounts lower than 1%wt were not improving the ionic conductivity, perhaps because the additive was not enough to act as a filler. In contrast, an excessive amount of additive (5%wt), was detrimental to the sample's ionic conductivity. Therefore, the amount of 1%wt resulted in the best performance, consistent with a recent report, which used UHS and indicated 1%wt of $Li_3N$ as the optimal amount if used as a filler.[28] A list of the acronyms of the prepared samples without and with additives is provided in Table 1.

## 2.2 Physical characterization

The relative density of the pellets was calculated as the ratio of the experimentally measured density and the LAGP theoretical density. The experimental density was estimated using Archimedes' method. Conventional and UHS pellets had a relative density of 80%, consistent with previously reported LAGP pellets prepared by sintering glass powders at 750°C.[42]

The crystal structure of the sintered materials was assessed by X-ray diffractometry (XRD, Bruker D8 ADVANCE, Germany) with Cu K-alpha radiation ($\lambda$=1.5406 Å). The cross-section of the fractured pellets was imaged by scanning electron microscopy (SEM, ZEISS Auriga equipped with a 30kV Gemini FESEM column, Germany). Thin lamellas of LAGP were prepared from LAGP pellets using a focused Ga-ion beam from an FEI Dual Beam Helios NanoLab 600 instrument. The thin lamellas were characterized by scanning transmission electron microscopy (STEM) and energy-dispersive X-ray spectroscopy (EDX) using a probe aberration-corrected FEI Titan Low-Base operating at 300 keV equipped with an Ultim Max sensor (Oxford Instruments, UK). Conventional and UHS samples were also characterized using a Raman spectrometer (Horiba Scientific, France) with a laser source wavelength of ~532 nm. DSC measurements from room temperature to 850°C at heating rates of 5, 10, 30, 50, and 100 °C/min were carried out to identify the glass-transition ($T_g$), crystallization onset ($T_c$), and crystallization peak ($T_p$) temperatures using a DSC 1 Star System apparatus (Mettler Toledo, USA, with a temperature precision of ±0.2 °C). For these measurements, ~15 mg of LAGP glass powders were loaded in a 70 µL alumina pan, and an empty alumina crucible was used as a reference.

## 2.3 Electrochemical characterization

Both sides of the LAGP pellets were sputtered with current collecting Au leads. EIS was carried out in air between 25 ºC and 65 ºC using a ProboStat test station (ProboStat™, NorECs, Norway). For those measurements, a Novocontrol Alpha-A high-performance impedance analyzer (Novocontrol technologies, Germany) was used in a 1Hz-10MHz frequency range. For EIS measurements at low temperature (-60 ºC), the samples were mounted on a Linkam stage (Linkam Scientific Instruments, UK) equipped with an LNP95 (Linkam Scientific Instruments, UK) liquid nitrogen pump. The EIS spectra were fitted using the Z-view® software and the equivalent circuit model shown and discussed in Section 3.3. The two blocks in series of resistance and constant-phase element (CPE) enabled the identification of the resistance and CPE of bulk ($R_{bulk}$, $CPE_{bulk}$) and grain boundary ($R_{gb}$, $CPE_{gb}$) at high and low frequency, respectively.

The total resistance was calculated as $R_{tot}=R_{bulk}+R_{gb}$, and correspondingly the total ionic conductivity, $\sigma_{tot}$, is defined as $\sigma_{tot}=\frac{1}{R_{tot}} \cdot \frac{t}{A}$, where t and A are the thickness and surface area of the samples, respectively. The samples' activation energy ($E_a$) for Li$^+$ conduction was determined according to the Arrhenius equation, $\sigma_{tot} T = B \cdot \exp(-\frac{E_a}{k_B T})$, where B is a pre-exponential factor, $k_B$ is the Boltzmann's constant, and T is the temperature. The $E_a$ can be estimated from the slope of the Arrhenius plot $\log(\sigma_{tot}T)$ *vs.* 1000/T using linear regression. The mean value and error (the error is given as the standard deviation estimated from linear regression) of $E_a$ are provided in the text when discussing the $Li^+$ conduction properties.

The conductivity of the bulk was calculated as $\sigma_{bulk} = \frac{1}{R_{bulk}} \cdot \frac{t}{A}$. The brick-layer model (BLM) was used to estimate the grain boundary conductivity.[44-48] The BLM assumes the conductor is an assembly of "bricks" characterized by a size D and grain boundary thickness $\delta$.[44, 48] As a result, the conductivity of the grain boundary was computed as $\sigma_{gb} = \frac{1}{R_{gb}} \cdot \frac{t}{A} \cdot \frac{\delta}{D}$. Assuming equal dielectric constant for bulk and grain boundary, as is generally assumed for LAGP and similar materials,[42, 49-50] it follows that

$$\frac{\delta}{D} = \frac{C_{bulk}}{C_{gb}} \qquad (1)$$

where $C_{bulk}$ and $C_{gb}$ are the capacitances of the bulk and grain boundary that were calculated as $C = Q^{1/n} \cdot R^{\frac{1}{n}-1}$ (Q and n are parameters of the CPE).[42] Thus, the specific conductivity of the grain boundary can be computed as

$$\sigma_{gb} = \frac{1}{R_{gb}} \cdot \frac{t}{A} \cdot \frac{C_{bulk}}{C_{gb}} \qquad (2)$$

Uncertainty in $\sigma_{bulk}$ and $\sigma_{gb}$ estimates was determined by parametric bootstrap.[51] The parameters $R_{bulk}$, $R_{gb}$, $CPE_{bulk}$, and $CPE_{gb}$, used for estimating $\sigma_{bulk}$ and $\sigma_{gb}$, were modeled as normal random variables where mean and error values were obtained with the Z-view® software. The mean and standard deviations of $\sigma_{bulk}$ and $\sigma_{gb}$ were estimated from 1000 independent samples of $R_{bulk}$, $R_{gb}$, $CPE_{bulk}$, and $CPE_{gb}$. The uncertainty in the pellets' geometry, *i.e.*, t and A, was discarded as those could be measured much more precisely than the electrochemical circuit parameters.

## 3  Results and Discussion

### 3.1  Sintering mechanism of ultra-fast annealed LAGP pellets

In glass, densification is determined by the viscous flow of glass particles,[52] which typically takes place above the glass-transition temperature, $T_g$. In contrast, a crystalline phase is formed at crystallization onset temperature, $T_c$. The quantity $T_c - T_g$ can be used to measure the stability of a glass towards crystallization, *i.e.*, the glass thermal stability window.[53] As LAGP glass systems typically have poor glass thermal stability window ($T_c - T_g$=80-90 °C) relative to other, more stable (for instance, $T_c - T_g$=200 °C for $Li_2O \cdot 2SiO_2$) glass compositions,[17, 54] crystallization and sintering are concurrent processes.[55] If these processes are concurrent, crystals can be formed during densification and hinder the viscous flow of the glassy phase and, therefore, the sintering process.

UHS enables high-temperature annealing with fast heating and cooling, which can influence the sintering and crystallization of glass-ceramics by shifting $T_g$, $T_c$, and the crystallization peak temperature, $T_p$.[33, 56] Therefore, DSC measurements with heating rates as high as 100 °C/min were performed to identify these temperatures. The measured DSC curves are shown in Figure 2 (a) and S3, where glass transition and crystallization are identified using the endothermic inflection and the exothermic peak, respectively. $T_g$, $T_c$, and $T_p$ for all the ramping rates (see Table S1) were estimated accordingly. As shown in Figure 2, $T_c$ and $T_p$ increase with increasing heating rates. In contrast, $T_g$ is far less sensitive to the heating rate (Figure S4), consistent with previous reports.[33]

Consequently, the glass thermal stability window of LAGP rises from 87 °C to 118 °C when the heating rate increases from 5 °C/min to 100 °C/min. As LAGP glass prepared by UHS is subjected to heating rates higher than 100 °C/min, crystallization is projected to occur at T>664 °C. Thanks to the delayed crystallization induced by UHS, sintering and crystallization of LAGP are no longer expected to be concurrent processes.

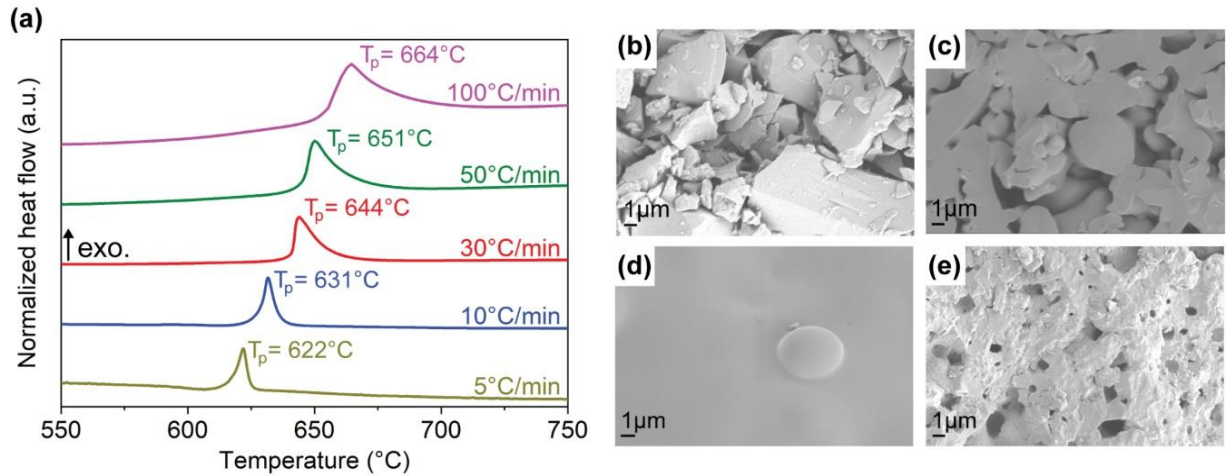

Figure 2. (a) DSC curves at various heating rates, showing that $T_p$ shifts towards higher temperatures if the heating rate increases. (b-e) Cross-section of the samples treated by UHS at different temperatures for 180 seconds: (b) T<<$T_g$ (T=400 °C), (c) T≈$T_g$ (T=550 °C), (d) $T_g$<T<$T_c$ (T=650 °C), and (e) T>$T_p$ (T=750 °C). Different stages of LAGP glass sintering are shown: (b) when T<<$T_g$, the sintering process is not triggered and the pellet consists of disjoint particles; (c) if T≈$T_g$, particles merge and sintering necks are visible; (d) if $T_g$<T<$T_c$, the material flows and a dense, amorphous monolith is obtained; (e) if T>$T_p$, crystallization occurs.

SEM cross-section images included in Figure 2 (b-e) show the pellet cross-sections treated for 180 seconds at different temperatures, *i.e.*, T<<$T_g$ (T=400 °C), T≈$T_g$ (T=550 °C), $T_g$<T<$T_c$ (T=650 °C), and T>$T_p$ (T=750 °C). Figure 2 (b) depicts the sample treated at T=400 °C; the pellet consists solely of pressed powders, as the temperature is too low for activating viscous flow and sintering. Conversely, in Figure 2 (c), sintering necks between particles are visible because at T=550 °C the sintering process is triggered together with the viscous flow of the glass particles.[52] For T=650 °C, see Figure 2 (d), a fully-dense amorphous microstructure is obtained as sintering continues without crystallization (see the XRD pattern in Figure S5 (a)) which is not attainable in conventional furnaces due to their low heating rates.[42] Lastly, when T=750 °C, the sample in Figure 2 (e) is entirely crystalline with no residual glassy phase being detected.

According to the microstructural evolution shown in Figure 2 (b-e), LAGP glass annealed by UHS can achieve complete sintering without crystallization, producing pellets with the amorphous XRD pattern and fully-dense microstructure in Figure 2 (d) and S4. Therefore, annealing by UHS can effectively decouple sintering and crystallization. Furthermore, it is noticeable that at this magnification level, no porosity can be observed in the dense-glassy pellet (Figure 2 (d)). In contrast, pores are visible only after crystallization (Figure 2 (e)). Therefore, porosity appears to result from density change amid crystallization due to the difference in density of the crystalline phase of LAGP compared to the parent glass.[54]

### 3.2 Crystal structure and microstructure of LAGP pellets

LAGP pellets were successfully annealed by UHS above the crystallization temperature (~750°C) in 180 seconds, see Figure 3 (a). Conventionally sintered LAGP pellets, requiring heat treatment for more than 16 hours, were also fabricated and characterized for comparison. XRD patterns of the conventional and UHS-sintered samples are presented in Figure 3 (b). All diffraction peaks can be indexed to the NASICON-type crystal structure of LAGP (JCPDS card #01-080-1924) with no detectable impurities. Instead, the XRD spectrum of the as-received amorphous powders has a broad peak of amorphous glass at 2θ=30° (the feature at lower 2θ was due to the sample holder). The absence of the 2θ=30° feature in the XRD spectra of the samples thermally treated above $T_c$ suggests complete crystallization. Raman spectra of LAGP conventional and LAGP UHS (Figure S6) are consistent with those reported in the literature for pure LAGP,[57-58] further supporting the absence of detectable impurities. In addition, despite the contact with carbon strips, LAGP UHS showed no evidence of graphite contamination, which was responsible for the blackening of crystalline LAGP treated by UHS.[13]

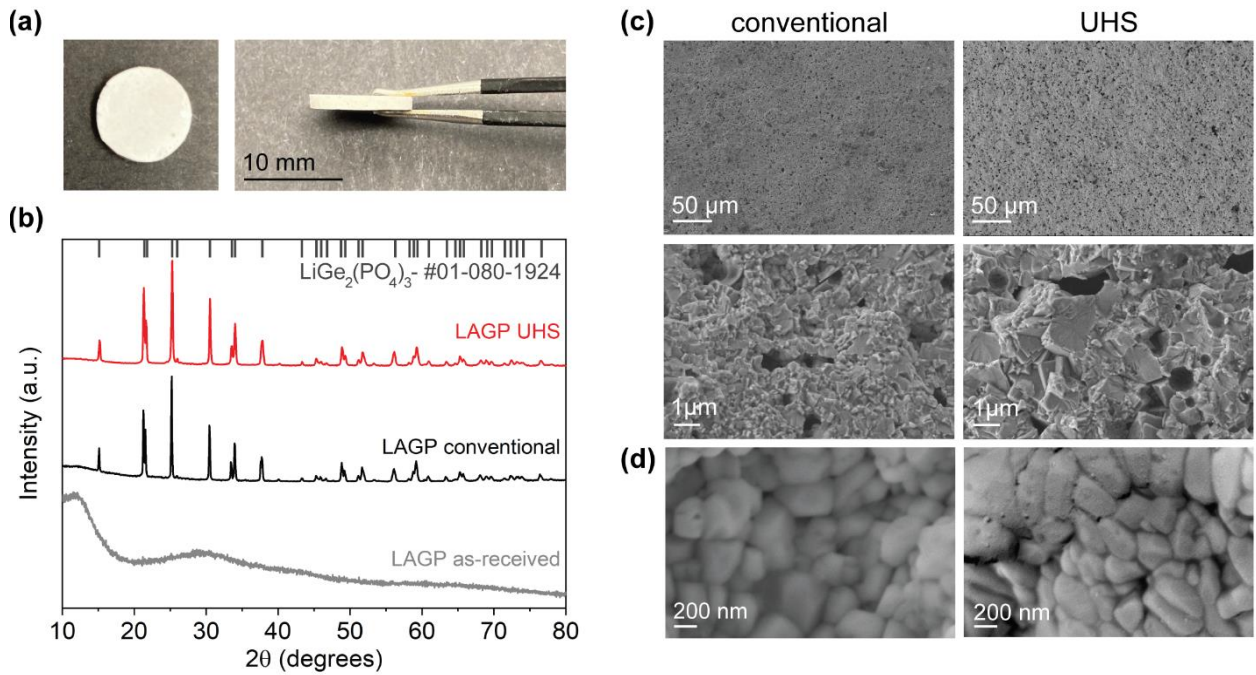

Figure 3. (a) Optical image of a pellet sintered by UHS; (b) XRD patterns of as-received LAGP glass and LAGP processed conventionally and by UHS. SEM micrographs of (c) the fractured cross-section and (d) the topmost free surface of LAGP pellets processed conventionally and by UHS.

The relative densities of the samples are 80% and 78% for the conventional and UHS pellets, respectively; the former value is consistent with the literature on glassy LAGP sintered at 750°C.[42] SEM micrographs of fractured cross-sections are displayed in Figure 3 (c). The estimated average grain size, see Figure 3 (d), was ~380 nm and ~260 nm for the conventional and UHS samples, respectively, as UHS's short processing time does not allow the nuclei of LAGP UHS to coarsen, and smaller grains were obtained.

LAGP conventional and UHS samples were also characterized by STEM-EDX to assess the chemical composition at the microstructural level. Representative images of grains and grain boundaries of LAGP conventional and LAGP UHS are presented in Figure 4 (a) and (b), respectively. Figure 4 (a) shows that the conventionally prepared LAGP pellet suffers from composition inhomogeneity, likely due to the precipitation of a detrimental Al-rich, Ge-poor phase, probably associated with Li loss due to the long processing times. In contrast, LAGP

UHS does not feature detectable impurities, see Figure 4 (b), confirming that UHS effectively suppresses the precipitation of secondary phases.[27]

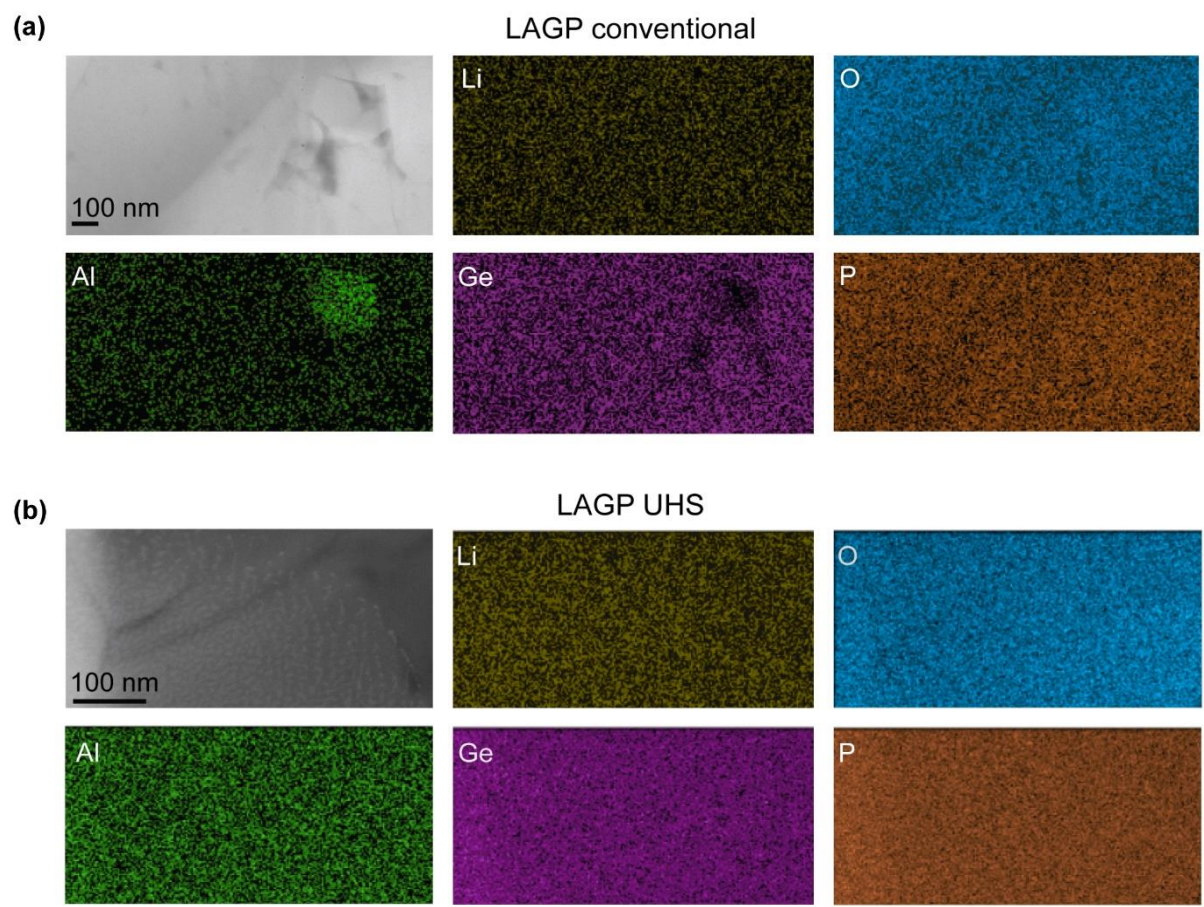

Figure 4. STEM and STEM-EDX micrographs of (a) LAGP conventional and (b) LAGP UHS. The investigated areas include both grain and grain boundaries. One can note in LAGP conventional the precipitation of an Al-rich, Ge-poor secondary phase.

### 3.3   Electrochemical characterization of LAGP pellets

Fabricated LAGP pellets were characterized by EIS to assess the impact of UHS on Li$^+$ transport. Amorphous LAGP is not an effective ion conductor.[42] Fully-dense glassy pellets obtained by UHS below $T_c$ (Figure 2 (d)) had extremely low ionic conductivity at room temperature, far lower than the value of $2\times10^{-7}$ S/cm estimated at 350 °C, see the EIS spectra in Figure S7.

In contrast, when LAGP crystallized into the NASICON-type structure, its Li$^+$ conductivity increased sharply. LAGP samples sintered and crystallized conventionally and by UHS were characterized by EIS at temperatures between 25 ºC and 65 ºC, as shown in Figure 5 (a) and (b). The collected EIS spectra are characterized by two semi-circles and a straight-line feature (see examples in Figures S8). The high (~$10^7$-$10^6$ Hz) and low frequency (~$10^6$-$10^3$ Hz) arcs are related to ion transport at bulk and grain boundary, respectively.[42, 49, 53, 59] To model these phenomena, the equivalent circuit model in Figure 5 (a) was used. The circuit included two blocks in series, each consisting of a resistor and a CPE, representing the resistance associated with Li$^+$ ion transport and the capacitance generated by dipole and double-layer formation at the bulk and grain boundary, respectively. Finally, a CPE was used to simulate charge accumulation in the ion-blocking gold electrodes.

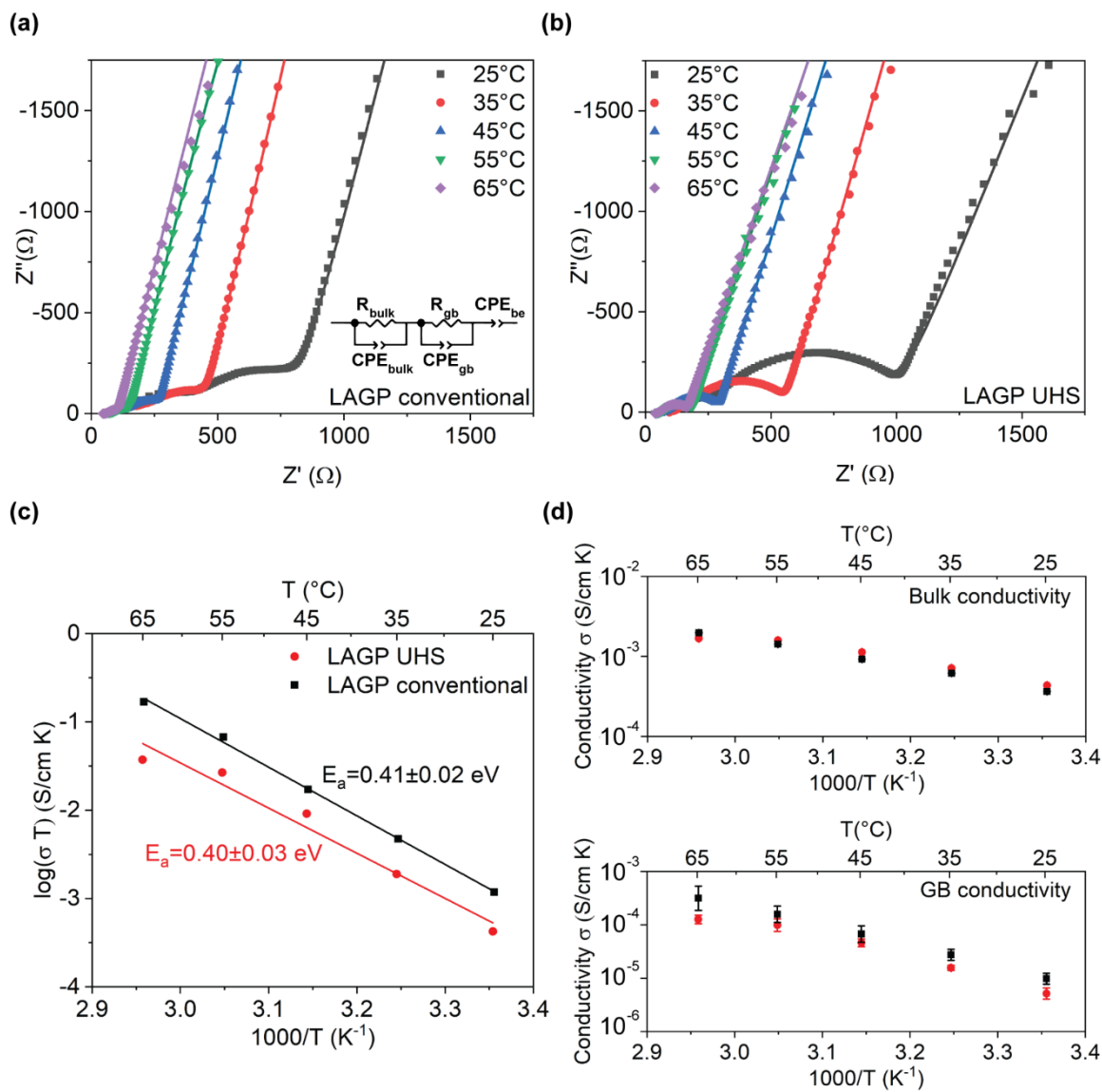

Figure 5. Experimental data and fitting of EIS spectra measured from 25°C up to 65°C for (a) LAGP conventional and (b) LAGP UHS. (c) Arrhenius plot of log(σT) *vs.* 1000/T and estimated activation energy for Li$^+$ transport ($E_a$) of LAGP conventional and LAGP UHS. (d) Bulk and grain boundary conductivities of LAGP conventional and LAGP UHS.

Arrhenius plots of the samples' ionic conductivity are shown in Figure 5 (c). Li-ion conductivity at room temperature and $E_a$ for the conventional and UHS samples are 1.75×10$^{-4}$ S/cm and 0.41±0.02 eV, and 1.15×10$^{-4}$ S/cm and 0.40±0.03 eV, respectively. Both pairs of conductivity and $E_a$ values are consistent with those reported for LAGP.[42] Despite the comparable ionic conductivity and activation energy for ion transport, the EIS spectra of LAGP conventional and LAGP UHS have significantly different characteristics, see Figure S8.

Specifically, the low-frequency arc, corresponding to Li$^+$ transport at the grain boundary, is substantially bigger for LAGP UHS than LAGP conventional. Such difference is even more evident in the EIS measurements at -60 °C (Figure S9), implying that LAGP UHS suffers from higher grain boundary resistance.

To shed light on the UHS sample's high grain boundary resistance, the BLM was used to estimate the ionic conductivity of the grain interior ($\sigma_{bulk}$) and grain boundary ($\sigma_{gb}$). As shown in Figure 5 (d), the bulk conductivities of conventional and UHS samples are similar (3.64×10$^{-4}$ S/cm *vs.* 4.30×10$^{-4}$ S/cm at room temperature, respectively). Conversely, conventional sintering and UHS yield different grain boundary conductivities. At room temperature, the specific conductivity of the grain boundary for LAGP conventional is 9.79×10$^{-6}$ ±2.07×10$^{-6}$ S/cm, while the one for LAGP UHS is ~40% lower, *i.e.*, 5.96×10$^{-6}$ ±1.28×10$^{-6}$ S/cm. Consequently, the difference in total ionic conductivity between the two samples can be ascribed to the lower grain boundary conductivity of the sample prepared by UHS.

Such a lower grain boundary conductivity may be attributed to high porosity, precipitation of insulating phases, or poor grain contact.[42] However, here, the first two hypotheses can be discarded, as LAGP conventional and LAGP UHS have similar densities, and XRD, Raman, and STEM-EDX did not detect impurities in the latter. Therefore, the lower grain boundary conductivity of LAGP UHS can be attributed to poor inter-grain contact, likely caused by the high heating rate and short processing time of UHS. Indeed, we must consider that as sintering and crystallization occur quickly during UHS, grain coarsening is hindered. To solve this issue, the use of two additives are studied in the next section.

## 3.4 Microstructural and electrochemical characterization of LAGP pellets fabricated with additives

As discussed above, LAGP prepared by UHS is characterized by lower grain boundary conductivity than the one prepared conventionally. To overcome this issue, we fabricated LAGP using the additives $B_2O_3$ and $Li_3BO_3$, which have been reported to improve grain contact and grain boundary conductivity. To assess the effect of these additives, we fabricated samples with compositions LAGP + 1%wt $B_2O_3$ (LAGP-B) and LAGP + 1%wt $Li_3BO_3$ (LAGP-LBO).

### 3.4.1 Effect of $B_2O_3$ addition

XRD patterns and SEM micrographs of conventionally and UHS annealed LAGP-B are compared to those of pure LAGP in Figure S11 (a) and (b). LAGP single phase can be indexed, *i.e.*, the addition of $B_2O_3$ to the amorphous LAGP powders did not trigger the formation of detectable impurities or secondary phases. In addition, the pellets were characterized by comparable relative density (~78%). The SEM micrographs of LAGP-B and LAGP UHS indicate that the two samples have similar microstructures. The EIS spectra in Figure S11 (d) suggest that the total resistance of LAGP-B is lower than that of pure LAGP, implying higher ionic conductivity ($1.97 \times 10^{-4}$ S/cm *vs.* $1.15 \times 10^{-4}$ S/cm at room temperature, respectively).

The Arrhenius plot in Figure 6 (a) shows that the value of activation energy for ion transport of LAGP-B conventional and LAGP-B UHS is $0.40 \pm 0.01$ eV *vs.* $0.36 \pm 0.02$ eV, respectively. In addition, the ionic conductivity of LAGP-B prepared by conventional sintering is lower than that of LAGP-B UHS (~$1.70 \times 10^{-4}$ S/cm *vs.* ~$1.97 \times 10^{-4}$ S/cm at room temperature, respectively). As previously done for pure LAGP, we estimated the bulk and grain boundary conductivity, see Figure 6 (b) and Table S2. The conductivity of the grain bulk is not significantly different compared to LAGP, which is reasonable as $B_2O_3$ is not added to the glass formula during the synthesis of the samples. Therefore, $B_2O_3$ is not incorporated in the NASICON structure, and the bulk of LAGP-B has the same composition as LAGP. In contrast, upon $B_2O_3$ addition, the grain boundary conductivity is enhanced, especially in the case of LAGP-B UHS, which is double that of LAGP UHS ($1.15\times10^{-5} \pm 2.75\times10^{-6}$ S/cm *vs.* $5.96\times10^{-6} \pm 1.28\times10^{-6}$ S/cm, respectively).

It is worth noting that the addition of $B_2O_3$ leads to no improvement if conventional sintering is used. The total conductivities and $E_a$ of LAGP and LAGP-B are similar (~$1.75\times10^{-4}$ S/cm and $0.41\pm0.01$ eV *vs.* ~$1.70\times10^{-4}$ S/cm and $0.40\pm0.01$ eV at room temperature, respectively). Due to the volatility of $B_2O_3$ at the sintering temperature (750 °C),[60] the lengthy conventional sintering (12 hours) is less effective compared to UHS when low-melting additives are used.[28]

Finally, comparing Figure S11 (e) to Figure S9, it is noticeable that the grain boundary arc of LAGP-B, despite being comparatively smaller than that of LAGP UHS, still dominates the EIS spectrum of LAGP-B UHS. Therefore, it may be possible that using a Li-based additive such as $Li_3BO_3$, which besides improving inter-grain contact, also introduces excess lithium, could further improve $Li^+$ conduction.[49]

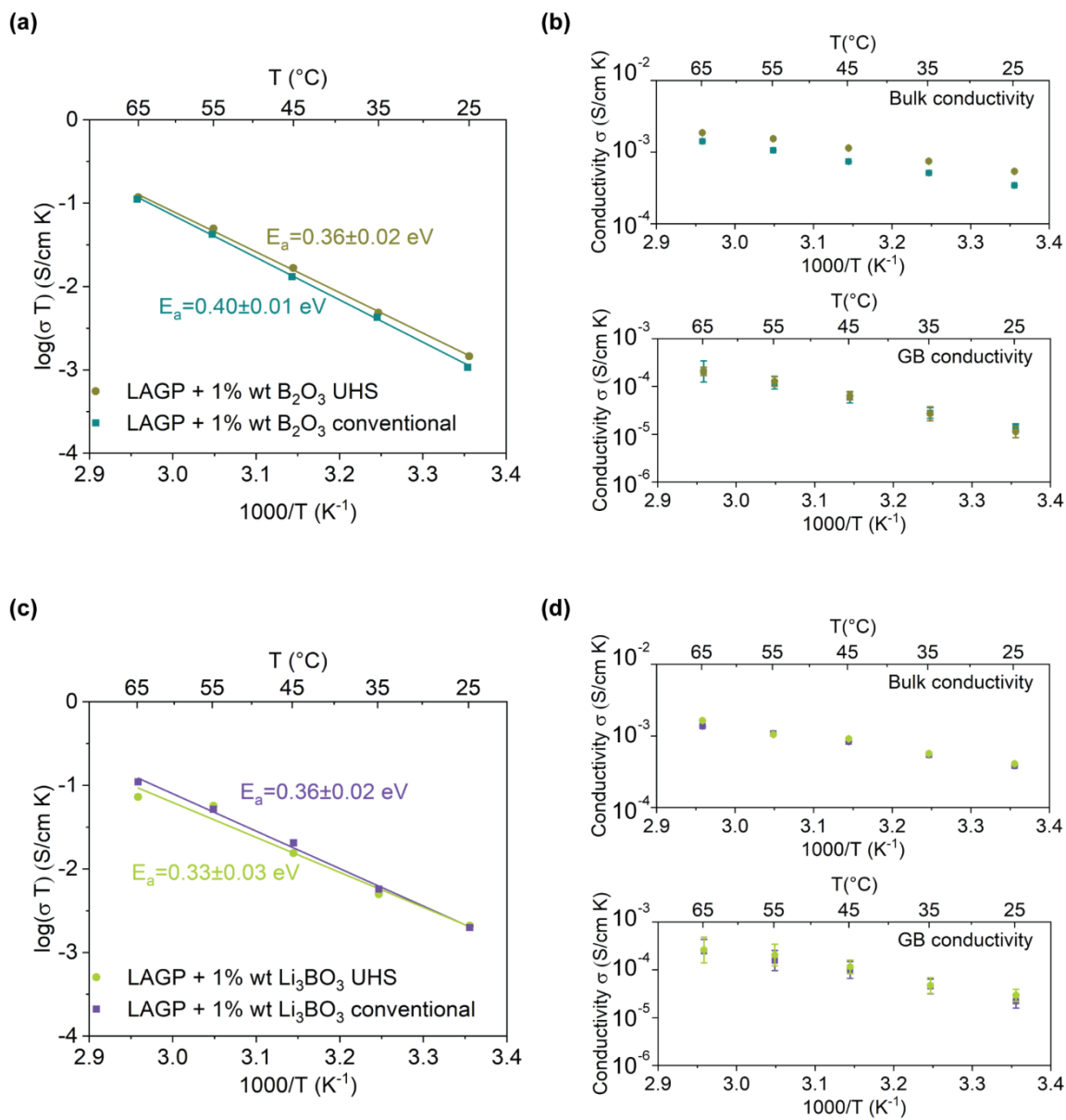

Figure 6. (a) Arrhenius plot of log(σT) *vs.* 1000/T and estimated $E_a$ for Li$^+$ transport of LAGP-B conventional and LAGP-B UHS. (b) Bulk and grain boundary conductivity of LAGP-B conventional and LAGP-B UHS. (c) Arrhenius plot of log(σT) *vs.* 1000/T and estimated $E_a$ for Li$^+$ transport of LAGP-LBO conventional and LAGP-LBO UHS. (d) Bulk and grain boundary conductivity of LAGP-LBO conventional and LAGP-LBO UHS.

### 3.4.2 Assessing the effect of Li$_3$BO$_3$ addition

XRD spectrum of the LAGP-LBO has three small peaks in addition to those of LAGP, see Figure S12 (a), which can be attributed to Li$_3$BO$_3$ (see XRD patterns at the bottom of Figure S12 (a)). The relative density of LAGP UHS and LAGP-LBO UHS is similar, about 78% and 75%, respectively. SEM micrographs in Figure S12 (b) show that the porosity in the LAGP-LBO sample is slightly higher, consistent with a previous study reporting a slight decrease of pellet's density upon addition of $Li_3BO_3$ to $LiTi_2(PO_4)_3$.[39]

LAGP-LBO reached an ionic conductivity of ~$2.25\times10^{-4}$ S/cm and ~$2.3\times10^{-4}$ S/cm when sintered by conventional and UHS methods, respectively. The activation energy for ion transport of the two samples is $0.36\pm0.02$ eV and $0.33\pm0.03$ eV, see Figure 6 (c), which is significantly reduced compared to LAGP. Further, the bulk and grain boundary conductivities were computed and plotted in Figure 6 (d). As $Li_3BO_3$ did not enter the NASICON structure, the bulk composition was not expected to change, and the bulk conductivity of LAGP-LBO samples was virtually identical to that of LAGP, see Table S2. In contrast, the specific conductivity of the grain boundary shows, as expected, a significant improvement for both samples. Remarkably, LAGP-LBO UHS reaches a grain boundary conductivity of $3.06\times10^{-5}\pm7.4\times10^{-6}$ S/cm, a value ~5 times higher than that of LAGP UHS, see Table S2. Finally, the EIS spectrum of LAGP-LBO UHS at a sub-zero temperature in Figure S12 (e) shows arcs with a similar radius for bulk and grain boundary, confirming that the latter is no longer the main contribution to the spectrum.

## 4    Conclusions

In this work, LAGP solid electrolytes were sintered and crystallized from glassy LAGP powders in 180 seconds by UHS. According to DSC experiments, fast heating rates delay the crystallization temperature of LAGP. The delayed crystallization induced by UHS results in a wider glass thermal stability window, potentially decoupling the sintering and crystallization phenomena and enabling fully-dense glassy pellets.

Further, EIS characterizations indicate that LAGP produced by UHS has slightly lower ionic conductivity than the one treated by conventional sintering ($1.15\times10^{-4}$ S/cm *vs.* $1.75\times10^{-4}$ S/cm, respectively). By deconvolving bulk and grain boundary contributions to the total resistance and applying the BLM, the lower conductivity of LAGP UHS can be attributed to hampered Li$^+$ transport at grain boundaries, likely due to poor inter-grain contact. To overcome this issue and enhance the low-conducting grain boundaries, the use of $B_2O_3$ and $Li_3BO_3$ additives was explored. As a result, LAGP annealed by UHS with either $B_2O_3$ or $Li_3BO_3$ possessed a grain boundary specific conductivity of $1.15\times10^{-5}$ S/cm and $3.06\times10^{-5}$ S/cm, respectively. These values are ~2 and ~5 times higher than that of LAGP UHS ($5.96\times10^{-6}$ S/cm).

Overall, this work represents a novel contribution toward understanding the influence of UHS on the physical and electrochemical properties of crystalline LAGP sintered from glassy LAGP powders. In addition, this work suggests that UHS annealing at ambient conditions is an attractive pathway for the scaled-up production of all-solid-state batteries.

Table 1. List of acronyms.

| | |
|---|---|
| **UHS** | Ultrafast high-temperature sintering |
| **LAGP** | $Li_{1.5}Al_{0.5}Ge_{1.5}(PO_4)_3$ |
| **BLM** | Brick-layer model |
| **EIS** | Electrochemical impedance spectroscopy |
| **XRD** | X-ray diffraction |
| **SEM** | Scanning electron microscopy |
| **STEM** | Scanning transmission electron microscopy |
| **EDX** | Energy-dispersive X-ray spectroscopy |
| **DSC** | Differential scanning calorimetry |
| **$T_g$** | Glass-transition temperature |
| **$T_c$** | Crystallization onset temperature |
| **$T_p$** | Crystallization peak temperature |
| **LAGP-B** | Electrolyte composed of LAGP + 1%wt $B_2O_3$ |
| **LAGP-LBO** | Electrolyte composed of LAGP + 1%wt $Li_2BO_3$ |

**Supporting Information**

Additional details about characterizations of physical and electrochemical properties with additional plots (SEM micrographs, XRD patterns, impedance spectroscopy measurements) and tables

**Acknowledgments**

The authors gratefully acknowledge the help of Raul Arenal and the Laboratory of Advanced Microscopy at the University of Zaragoza for the STEM-EDX characterizations. A. Curcio kindly recognizes the support of the Hong Kong Ph.D. Fellowship Scheme. A. Curcio and F. Ciucci gratefully acknowledge the financial support from the Research Grants Council of Hong Kong (RGC Ref No. 16201820 and 16206019). A. G. Sabato acknowledges that this project has received funding from the European Union's Horizon 2020 research and innovation program under the Marie Skłodowska-Curie Grant Agreement No. 841937. J. C. Gonzalez-Rosillo acknowledges the financial support provided by the European Union's Horizon 2020 research and innovation program under the Marie Skłodowska-Curie Grant Agreement No. 801342 (Tecniospring INDUSTRY), as well as by the Agency for Business Competitiveness of the Government of Catalonia. This research was partially funded by the Generalitat de Catalunya (2017 SGR 1421, NANOEN) and the Spanish Ministry of Science and Innovation under the Grant PID2019-107106RB-C31 (RETOS, 3DPROGRESS). This work was supported in part by the Project of Hetao Shenzhen-Hong Kong Science and Technology Innovation Cooperation Zone (HZQB-KCZYB-2020083). The authors also thank the Materials Characterization and Preparation Facilities (MCPF) of HKUST for their kind technical assistance.